\input psfig
\documentstyle[12pt,titlepage]{article}

\newcommand{\cen}[1]{\centerline{#1}}

\begin{document}

\newcommand{\su}{\hspace*{.1in}}
\newcommand{\para}{\par}
\newcommand{\be}{\begin{equation}}
\newcommand{\en}{\end{equation}}
\newcommand{\sot}{\hspace*{.8in}}

\newcommand{\loe}{\stackrel{<}{\sim}}
\newcommand{\goe}{\stackrel{>}{\sim}}

\newcommand{\ggg}{$\gamma$}
\newcommand{\eee}{$e^{\pm}$}
\newcommand{\lap}{$L_{38}^{-1/3}$}
\newcommand{\ergs}{\rm \su  erg \su s^{-1}}
\newcommand{\etal}{ {\it et al.}}

\def\jref#1 #2 #3 #4 {{\par\noindent \hangindent=3em \hangafter=1
      \advance \rightskip by 0em #1, {\it#2}, {\bf#3}, #4.\par}}
\def\rref#1{{\par\noindent \hangindent=3em \hangafter=1
      \advance \rightskip by 0em #1.\par}}

\def\p{\phantom{1}}
\def\pmu{\mox{$^{-1}$}}
\def\ApJ{{\it Ap.\,J.\/}}
\def\ApJL{{\it Ap.\,J.\ (Letters)\/}}
\def\ApJS{{\it Astrophys.\,J.\ Supp.\/}}
\def\AJ{{\it Astron.\,J.\/}}
\def\AAL{{\it Astr.\,Astrophys.\ Letters\/}}
\def\AAS{{\it Astr.\,Astrophys.\ Suppl.\,Ser.\/}}
\def\MN{{\it Mon.\,Not.\,R.\,Astr.\,Soc.\/}}
\def\Na{{\it Nature \/}}
\def\SAIt{{\it Mem.\,Soc.\,Astron.\,It.\/}}
\def\BGD{\begin{description}}
\def\EDD{\end{description}}
\def\BGF{\begin{figure}}
\def\EDF{\end{figure}}
\def\BGC{\begin{center}}
\def\EDC{\end{center}}
\def\BGT{\begin{tabular}}
\def\EDT{\end{tabular}}
\def\BGE{\begin{equation}}
\def\EDE{\end{equation}}
\def\REFFF{\par\noindent\hangindent 20pt}
\def\DS{\displaystyle}
\def\kms{km^s$^{-1}$}
\def\sbu{mag^arcsec${{-2}$}}
\def\e{\mbox{e}}
\def\dex{\mbox{dex}}
\def\L{\mbox{${\cal L}$}}

\newcommand{\porb}{ P_{orb} }
\newcommand{\Po}{$ P_{orb} \su$}
\newcommand{\pdot}{$ \dot{P}_{orb} \,$}
\newcommand{\pot}{$ \dot{P}_{orb} / P_{orb} \su $}
\newcommand{\s}{ \\ [.15in] }
\newcommand{\mm}{$ \dot{m}$ }
\newcommand{\mdot}{$ |\dot{m}|_{rad}$ }
\newcommand{\myr}{ \su M_{\odot} \su \rm yr^{-1}}
\newcommand{\msol}{\, M_{\odot}}
\newcommand{\ppp}{ \dot{P}_{-20} }
\newcommand{\ci}[1]{\cite{#1}}
\newcommand{\bb}[1]{\bibitem{#1}}
\newcommand{\ch}[1]{\vskip .3in \noindent {\bf #1} \para}
\newcommand{\cms}{ \rm \, cm^{-2} \, s^{-1} }
\newcommand{\nn}{\noindent}

\renewcommand{\thebibliography}[1]{
  \list
  {[\arabic{enumi}]}{\settowidth\labelwidth{[#1]}\leftmargin\labelwidth
    \advance\leftmargin\labelsep
    \usecounter{enumi}}
    \def\newblock{\hskip .11em plus .33em minus .07em}
    \parsep -2pt
    \itemsep \parsep
    \sloppy\clubpenalty4000\widowpenalty4000
    \sfcode`\.=1000\relax}

\newcommand{\asca}{{\it ASCA} }

\def\psr{PSR~B1259-63 }
\def\psrp{PSR~B1259-63}
\def\ergs{{\rm \; erg \, s^{-1}}}
\def\ros{{\sl ROSAT }}
\def\grad{$^\circ$}
\def\asec{\ifmmode ^{\prime\prime}\else$^{\prime\prime}$\fi}
\def\amin{\ifmmode ^{\prime}\else$^{\prime}$\fi}
\def\fss{\hbox{$.\!\!^{\rm s}$}}        % Fractions of seconds
\def\fdg{\hbox{$.\!\!^\circ$}}           % Fractions of degrees
\def\msun{$M_{\odot}$}
\def\mdot{\dot M}
\def\grad{$^\circ$}
\def\fd{\hbox{$.\!\!^{\rm d}$}}            % Fractions of days

\baselineskip 19pt

\vspace*{.1in}

\cen{\Large \bf X-ray emission from the  PSR B1259--63 system}
\cen{\Large \bf  near apastron}
\vskip .3in
\cen{\large J. Greiner}
\cen{\it Max-Planck-Institut f\"ur Extraterrestrische Physik,
             85740 Garching, Germany}

\vskip .15in
\cen{\large M. Tavani\footnote{
New address: Columbia Astrophysics Laboratory, Pupin Hall,
Columbia University, New York, NY 10027.}}
\cen{\it Joseph Henry Laboratories and Physics Department, Princeton
             University, Princeton, NJ 08544}
\vskip .15in
\cen{\large T. Belloni}
\cen{\it Astronomical Institute "A. Pannekoek", University of Amsterdam}
\cen{\it  Kruislaan 403, 1098 SJ Amsterdam, The Netherlands}

\vskip .3in
\cen{Submitted to the {\it Astrophysical Journal Letters}: September 23, 1994}
\cen{Revised: December 2, 1994}

\vspace*{.4in}

\baselineskip 17pt
\centerline{\bf Abstract}

 The PSR B1259--63 system contains a 47 ms radio pulsar in a highly eccentric
 binary with a Be-star companion. Strongly time variable X-ray emission was
 reported from this system as the pulsar was near apastron in 1992-early 1993.
 The variability was primarily deduced from an apparent non-detection of the
 \psr system during a first pre-apastron \ros observation in February~1992.
 We have re-analyzed the \ros observations of the \psr system. Contrary to the
 results of a previous analysis, we find that the \psr system was detected by
 \ros  during the first off-axis February~1992 observation. The  intensity of
 the soft X-ray emission of the \psr system  before and after the 1992 apastron
 appears to vary at most by a factor $\sim 2$. Our results sensibly constrain
 theoretical models of X-ray emission from the \psr system.

\vskip .2in
\baselineskip 15pt
\nn
\rref{{\it Subject headings}:
binaries: close -- pulsars: individual (PSR~1259-63)
-- stars: emission-line, Be -- stars: mass loss -- X-rays: stars}

\baselineskip 20pt
\newpage
\vskip .2in
\centerline{\bf Introduction}
\vskip .1in

The 47.7 ms radio pulsar \psr  was discovered in 1991 during  a
1520 MHz radio survey of the southern galactic plane (Johnston \etal 1992a).
A study of timing residuals revealed that the pulsar was  in a very eccentric
binary ($e \sim 0.87$), large orbital period $P_{orb} \sim 3.4$ yrs,
and large mass function. The massive companion was
subsequently optically identified with the Be-star SS 2883 (= LS 2883 =
CPD--63 2495 = He 3-852) (Johnston \etal 1992b).  %mt101992b).
The pulsar/Be-star binary is near the galactic plane and its estimated
distance is 1.5--2.5~kpc (Johnston \etal  1992b; 1994). The \psr system is
particularly interesting for many reasons. The companion star is a massive
Be-star (with probable mass $m \sim$ 10 \msun) and   this system  is
a likely progenitor of  high-mass X-ray binaries. Furthermore, it is the only
known binary containing a rapidly rotating  radio pulsar and a Be-star.
Since Be-stars are characterized by large equatorial mass outflows,
the \psr system is ideal for the study of Be star outflows and of the
interaction of pulsar relativistic winds with gaseous nebulae (Tavani, 1994).

The \psr system  was  observed by the \ros PSPC instrument  three times
during the period February~1992 - February~1993 (see Table~1 for a
log of \ros observations). The first \ros observation was carried out before
apastron (apastron date: $\sim$ May~1,~1992) and two more observations were
carried out after apastron. For a companion mass $m=10 \, M_{\odot}$, the
apastron distance $a$  between the pulsar and the Be-star is  $a \sim
10^{14}$~cm corresponding to $a/R_{\star} \sim 350-250$, for a  Be-star
radius $R_{\star} = (6-10) \, R_{\odot}$. The source was clearly
 detected during the second and third  PSPC pointed observations
in early September 1992 (at $\phi \sim 5^{\circ}$ {\it after} apastron)  and
in February 1993 (at $\phi \sim 13^{\circ}$ {\it after} apastron)
 (Cominsky \etal 1994; hereafter CRJ94). The PSPC count rates for these
observations were reported to be  between 0.024 and 0.035 cts/s (see Table~1)
(CRJ94) and the implied  soft X-ray luminosity (0.1-2~keV) to be in the range
$L_x \sim (1-10)\cdot 10^{33} \ergs$ depending on the assumed spectrum (CRJ94).

The  early \ros PSPC observation of \psr (at  $\phi \sim$ 2\fdg7 {\it before}
apastron) was carried out in February~1992 with a  38 arcmin off-axis angle.
A first study of the data of this early \ros observation was carried out
by Bailes \& Watson (B\&W) and the results  are quoted in CRJ94. It was
reported that  the  early pre-apastron \ros observation failed to detect the
\psr system. The derived  upper limit on the count rate for the pre-apastron
observation was about a factor of ten lower than
the count rate measured during the  later post-apastron \ros observations.
The implied  asymmetry of the  X-ray emission between the apastron-approaching
and the apastron-receding parts of the orbit is potentially very important
for the interpretation of the nature of the X-ray emission. The apparent
X-ray asymmetry of \psr near apastron prompted an  interpretation in terms of
 a velocity-dependent interaction of the pulsar with  the  Be-star outflow
(King \& Cominsky, 1994; hereafter KC94).

Motivated by the importance of a possible pre- and post-apastron asymmetry of
the X-ray emission, we have re-analysed all \ros observations of \psr which
are now publicly available. In this paper we focus our discussion on the
detection by \ros  of the \psr system before  apastron. In the next section we
describe  the analysis and the results which substantially differ from  the
results previously  quoted in the literature. We briefly comment on the
implications of our results in the last section of the paper.

\vskip .3in
\centerline{\bf Re-analysis of \ros observations of the \psr system}
\vskip .1in

The EXSAS package (Zimmermann et al. 1993) has been used for the data
reduction.
The adapted source detection technique consists of several steps:
first, a mask is created to screen all the parts of the image where the
support structure of the PSPC entrance window may affect the
detectability of X-ray photons. Then, all possible sources are
identified by means of a "sliding window" technique and removed from the
data. A background map is produced with a bi-cubic spline fit to the
resulting image. Finally, a maximum likelihood algorithm is applied
to the data (e.g.,  Cruddace, Hasinger \& Schmitt 1988) in three separate
PHA channel ranges. If a  source is detected in more than one energy band,
a detection corresponding to a higher likelihood  value  is considered.

We apply this technique to the first \ros observation (February 1992)
of the \psr system. For the whole PSPC field of view we detect a total of
20 sources. One of these sources is at the position
$\alpha$=13$^h$02$^m$54\fss7, $\delta$=--63\grad49\amin06\asec\ (2000.0)
which is roughly 1 arcmin away from  the nominal (radio) position of
PSR~B1259--63 (Johnston \etal, 1994). In the following, we refer to this
source as `source-X'. We detect a total of $N_{\gamma} \sim 67$ photons from
the  source-X  corresponding to a likelihood of detection ${\cal L} \sim  27$.
Fig.~1 shows the \ros field near source-X with the pulsar position
 marked by a cross.

%______________________________________________ Gamma_1 (lg rho, lg e)
   \begin{figure}[thbp]
      \centering{
      \hspace*{-.5cm}
       \vbox{\psfig{figure=psr1259_ima.ps,width=10cm,%
          bbllx=7.2cm,bblly=12.82cm,bburx=18.8cm,bbury=24.cm,clip=}}\par
      }
      \caption[image]{Western part of the  \ros PSPC
              image of the February ~1992 observation of the field
              around the PSR~B1259--63 system. The cross
               shows the  position of the radio pulsar PSR~B1259-63
    [$\alpha$=13$^h$02$^m$47\fss72, $\delta$=--63\grad50\amin08\asec\ (2000.0)
     Johnston \etal, 1994]. The  X-ray source (`source-X') which is
              almost coincident with the position of PSR~B1259-63,
              appears  elongated and blurred due to the degraded point spread
              function of the \ros X-ray mirror for large off-axis angles.
              The bright object in the lower left corner is the 9.5 mag
              star HD 113466. The circular arc is the edge of the \ros
              field of view. The scale of 5~arcminutes is  indicated
              in the lower part of the figure.
              }
      \label{ima}
    \end{figure}

Since source-X  is  38 arcmin off-axis in the \ros field of view, its
appearance is distorted by the relatively large and asymmetric point spread
function of the \ros X-ray mirror for large off-axis angles. From  previous
off-axis \ros observations of bright X-ray sources, it is known that the best
fit centroid position of a gaussian approximation  to the asymmetric point
spread function introduces a systematic error which increases  with
increasing off-axis angle. Therefore, we estimate that the  apparent
positional difference between source-X and the radio pulsar is within the
statistical uncertainty of a typical off-axis observation. If, on the
contrary, source-X were not related to the pulsar system, our results would
indicate the existence of another X-ray source 1 arcmin apart from the
pulsar's position. However, this additional source has neither been reported
by CRJ94 nor found by our re-analysis of the on-axis pointed \ros PSPC
observations. The intensity decrease of this hypothetical additional X-ray
source would have been   at least by a  factor of $\sim$40 within a half year
interval. Although we cannot entirely exclude  this possibility, it is more
natural to relate the February 1992 source-X  to the \psr system\footnote{
The upper limit to the February~1992 ROSAT PSPC flux is now known to be
erroneous  due to an incorrect source position used to determine the source
flux from the \psr system. A new determination made independently by
Watson is entirely consistent with the value quoted here. (M.G. Watson,
private communication.)}. The  count rate from the \psr system of the
February~1992 observation  is ${\cal C} \sim 0.014 \pm 0.003$~cts/s, which
is lower by about 30\% than the value of the subsequent post-apastron
observation in September~1992. Table~1 gives a summary of  all \ros
observations of the \psr system.  Columns (3) and (4) give  the values
reported   in CRJ94. The CRJ94 analysis was carried out  only for the
September~1992 and February~1993 observations.

 \begin{table}[cr]
   \begin{center}
   \caption{Summary of the  \ros observations of the \psr system}
   \label{cr}
   \begin{tabular}{|lc|c|cc|cc|}
   \hline
   Observation & Date & Phase$^{\dag}$ &\multicolumn{2}{c|}{%CRJ94
                                                           previous values} &
   \multicolumn{2}{c|}{values of our re-analysis} \\
   \hline
          &      &(degrees)& source   & PSPC count  & source   & PSPC count  \\
          &      & & photons  & rate (cts/s) & photons  & rate (cts/s) \\
       \hline
B\&W   & Feb. 25-26 1992 & -2.7 &$<$ 10 & $<$ 0.0026  & 67 & 0.014$\pm$0.003 \\
CRJ94 obs.1 & Aug.30-Sep.4,1992& 5.&254 & 0.024$\pm$0.0018 & 204 &
0.020$\pm$0.002 \\
CRJ94 obs.2    & Feb.7-16, 1993 &13.& 1136 & 0.031$\pm$0.0011&1007&
0.029$\pm$0.001\\
       \hline
       \end{tabular}
       \end{center}
       \nn
       $\dag$ Orbital phase (in degrees) with zero set at apastron and
        negative (positive) values for apastron-approaching (receding) pulsar
        positions in the orbit.
   \end{table}

Our results  imply  that the soft X-ray flux of the \psr system  is only
marginally time-variable (within  $\sim$30\%) between  the pre-apastron
observation in February~1992 and  the  first post-apastron observation in
September~1992. Between February~1992 and February~1993 the soft X-ray flux
of the \psr system appears to have increased by a factor $\sim 2$. This
moderate increase of detected X-ray  flux during a 1-year period is probably
intrinsic to the \psr system and partly %mt10 due to caused by instrumental
effects.

Due to the limited number of photons detected in the February~1992 observation,
a spectral fit of the \ros PSPC data cannot be performed without additional
assumptions. As a first step in comparing the X-ray spectrum of the
February~1992  vs. the September~1992, we studied  the binned count rate
spectra without any fitting. Within the uncertainties, the two spectra
appear to be  identical. As a second step, in order to fit the absorbing
column density $N_H$, we have adopted a single power law model with
 photon index chosen to be either 1.9 or 1.6 [taken from ASCA observations
near periastron (Tavani \etal, 1994b; Kaspi et al., 1994)  and at $\phi \sim
125^{\circ}$ after periastron (Hirayama \etal, 1995)]. Again, our  results
are consistent with both X-ray spectra having the same absorbing column density
 $N_H = (2.5-8)\times 10^{21} \rm \,  cm^{-2}$. If we denote by ${\cal A}$ the
ratio of column densities  N$_H$ for the February~1992 and for the
September~1992 observations, we find that ${\cal A} \simeq 1$ within the
uncertainties of the spectral analysis of these PSPC data. While a value of
${\cal A} \loe 3$ is marginally possible, any value of ${\cal A} \goe 4-5 $
is certainly excluded. The  upper limit  on ${\cal A}$ is estimated by a
conservative assumption about the PSPC spectral capability and taking into
account the low number of photons detected in the two pointings.

We also performed a timing analysis of the X-ray data from the \psr
searching for periodic pulsations with the radio pulsar spin period
Near apastron the orbital correction due to the Doppler shift is
negligible and the nominal pulsar spin period can be obtained from the pulsar
ephemeris of Johnston \etal (1994).
We do not detect any X-ray  pulsation with the pulsar spin period (CRJ94).
 We used  standard X-ray timing techniques (e.g., Leahy et~al. 1983),
with  $n=5$ phase bins and an assumed sinuoidal pulsed signal.
The 90\% confidence limits on the pulsed fraction are
$f \loe 25$\% for the September~1992 observation ($N_{\gamma} \sim 200$)
and $f \loe 13$\% for the February~1993 observation ($N_{\gamma} \sim 1000$).

\vskip .3in
\centerline{\bf Discussion}
\vskip .1in

A possible source  of the observed X-ray emission of the \psr system is
coronal emission of the Be-star companion (also considered in CRJ94).
The likelihood of the interpretation   depends on the spectral classification
of the SS~2883 star and on its  distance. Since  no specific  spectral type
determination is available, Johnston \etal (1994) derive it indirectly  from
the colours of the SS~2883 star (in the range O9 -- B2). The system distance
is also uncertain. By assuming a main-sequence star of luminosity class V,
 the distance range of SS~2883 is $d\sim0.6-1.6$~kpc (Johnston \etal, 1994).
However, a pulsar dispersion measure model gives a larger estimate of the
distance, $d \sim 3-4$~kpc   (Taylor \& Cordes, 1993). The system distance is
most likely $d \sim 2$~kpc; we note that this is approximately the distance of
the first galactic  spiral arm in the  direction of \psrp.

A recent \ros X-ray survey of 12 nearby near-main-sequence B stars shows that
the ratio $\cal R$ of X-ray to bolometric luminosity is a sensitive function
of spectral type, ranging from ${\cal R} \sim 10^{-6}$ for O9 stars down to
${\cal R} \loe 10^{-8}$ for B2 stars (Cassinelli \etal 1994).
We notice that even for an adopted maximum ratio ${\cal R} \sim 10^{-6}$  and
distance $d \goe 1.5$~kpc, the  early-type star SS~2883 is expected to have
an  X-ray  luminosity $L \sim (2-3)\cdot 10^{32} \ergs$, i.e., a luminosity
lower by a factor $\sim 10$ than the estimated X-ray luminosity near apastron.
We are therefore left with two possibilities: (1)  either an early spectral
type SS~2883 star  produces the observed X-ray luminosity being at a distance
$d \ll 1.5$~kpc; or (2) the \psr distance is $d \goe $1.5~kpc and the observed
X-ray flux cannot be  produced by the  SS~2883  star (CRJ94).

For a more likely distance $d \goe 1.5$~kpc, other X-ray   emission mechanisms
need to be considered (Kochanek, 1993; CRJ94). Accretion of material onto the
surface of the neutron star is believed to quench the radio pulsar
mechanism and the pulsar relativistic wind. Since pulsed radio emission from
\psr was visible near the 1992 apastron (Johnston \etal, 1992b), any gaseous
material from the Be star companion cannot penetrate the pulsar light-cylinder
radius $R_{lc} = 2 \, \pi \, c/P \sim 2.8\cdot 10^8$~cm, with $c$ the speed
of light and $P=47$~ms the pulsar spin period. Accretion onto the surface of
the neutron star is therefore clearly inhibited near  the  1992 apastron of
the \psr system. The observed X-ray emission can be the result of
gravitational capture of gaseous material  near the pulsar light-cylinder
radius  (KC94).Alternately, unpulsed X-ray emission of luminosity
$L_x \sim 10^{33} \ergs$ near the periastron of \psr can be radiated by
shocked relativistic particles of the pulsar wind stopped at a
a shock radius $R_s \sim 10^{12} - 10^{13}$~cm$\, \gg R_{lc}$
(Tavani, Arons, Kaspi, 1994a, hereafter TAK94). In this `pulsar-driven'
mechanism of X-ray emission,  pressure balance is established  at $R_s$
between the  pulsar radiation pressure and the ram pressure of the Be star
outflow (TAK94).

 These two mechanisms differ in the dependence of the X-ray emission on the
relative velocity ${\bf {v}}_{rel}$ between the pulsar and the Be-star
outflow. The KC94  model, which assumes that  the pulsar radiation pressure
plays no role in the gravitational capture process,  depends critically on
${\bf {v}}_{rel}$. This model  can  explain a possible asymmetry of the
X-ray emission as the pulsar approaches to and recedes from apastron.
An  X-ray asymmetry can be produced  as the radial and/or tangential
components of ${\bf {v}}_{rel}$ become small for an assumed low-velocity
Be-star outflow (KC94). In order to  produce any significant asymmetric effect
on the mass captured by the pulsar gravitational field, the velocity of
the Be-star outflow near apastron must have a  small value, with radial and
tangential components near $ v_w \sim (10-30) \, \rm km \, s^{-1}$ (KC94).
The gravitational capture model predicts a variation of the ratio ${\cal A}$
 of absorbing column densities before and after apastron. The value of
${\cal A}$ is calculated to be in the range $4 \loe {\cal A} \loe 15$
 for $ v_w \sim (10-15) \, \rm  km \, s^{-1}$,
and ${\cal A} \sim 3$ for $ v_w \sim 30  \, \rm  km \, s^{-1}$ (KC94).
The results of  our analysis   constrain the gravitational capture model of
X-ray emission from the \psr system near apastron. An apastron outflow
velocity  slightly larger than the escape velocity from the binary,
$v_w \goe 30  \, \rm  km \, s^{-1}$, can be consistent with the observed level
of X-ray emission and with the upper limit of the allowed  range of ${\cal A}$
(A.~King \& L.~Cominsky, personal communication). There is no need  to invoke
a cool, very slow wind ($v_w \loe 20 \, \rm  km \, s^{-1}$) at large
distance from the Be~star.

On the other hand, the pulsar-driven  model based on standard  features of
Be star outflows (Waters \etal, 1988) predicts a shock  X-ray emission near
periastron which is marginally dependent on the temporal behavior of
 ${\bf {v}}_{rel}$.  For a constant Be-star outflow rate,  approximately
constant values of the  X-ray flux and of $N_H$  near apastron are  expected
(TAK94). A time-variable X-ray emission near apastron might  be caused by  a
change of  shock emissivity due to time variation of the shock radius. The
shock radius near apastron can change as  a consequence of  a time-variable
mass outflow rate from the Be-star companion.

\vskip .3in
\centerline{\bf Conclusions}
\vskip .1in

We have  re-analized the  \ros observations of the \psr system near apastron.
We show that  the \psr system was detected by \ros  also during  the first
\ros PSPC off-axis  observation on February 1992 when the  pulsar's orbital
phase was $\phi$ = 2\fdg7 before apastron. By comparing the first and
second \ros observations of the \psr system and taking into account the
different quality of off-axis vs. pointed \ros observations, we conclude that
the X-ray emission from the \psr system was approximately constant (within
30\%)  near the 1992 apastron. We find evidence for a possible increase
(within a factor of $\loe 2$) of the post-apastron X-ray emission in
early~1993.

We also obtained  the value of the absorbing column density before and after
apastron. We find that the value of the $N_H$ ratio  ${\cal A}$ is consistent
with being  approximately constant for the PSPC observations of the \psr
system studied here. A large  value of  ${\cal A} \goe 4-5$ is excluded
and the allowed range $1 \loe {\it A} \loe 3$ constraints the gravitational
capture model of X-ray emission (KC94).

X-ray observations of the \psr system near apastron are important
in determining the nature of the high energy emission in a regime
when  accretion onto the surface of the neutron star is  inhibited.
Future X-ray  observations of the \psr system near apastron to be carried out
by \ros and \asca  will be valuable in monitoring possible orbit-to-orbit
variations of the X-ray emission which can be caused by a time variable Be
star outflow rate. Spectral information and a determination of ${\cal A}$
will be available from \asca  observations  to be obtained in coincidence with
the 1995 apastron of the \psr system. These observations will provide crucial
information in determining the nature and properties of the shock emission
near apastron.

\vspace{.5cm}
\noindent
{\bf Acknowledgements}\\
We thank L. Cominsky and M. Roberts for useful discussions and correspondence,
which greatly improved this paper. We thank M.~Bailes, M.~Watson and A.~King
for a helpful exchange of information, and the referee for his comments.
MT also  thanks  V. Kaspi and F. Nagase for extensive discussions
and collaborative work  on the \asca periastron observations of \psrp,
and J. Arons for discussions. JG is supported by the Deutsche Agentur f\"ur
Raumfahrtangelegenheiten (DARA) GmbH under contract numbers FKZ~50~OR~9201.
MT acknowledges partial support by the  NASA grant NAG5-2593.
TB is supported by EC contract N. ERB-CHRX-CT93-0329 and by the Netherlands
Organization for Scientific Research (NWO) under grant PGS~78-277.

\newpage
\vskip .3in
\noindent
\centerline{\bf References}
\vskip .1in

 \rref{Cassinelli J.P., Cohen D.H., MacFarlane J.J., Sanders W.T.,
      Welsh B.Y., 1994, ApJ, 421, 705}

 \rref{Cominsky L., Roberts M., Johnston S., 1994, ApJ 427, 978 (CRJ94)}

\rref{Cruddace, R.G., Hasinger, G.R. \& Schmitt, J.H.M.M., 1988,
in {\it Astronomy from Large Databases}, eds. F. Murtagh F. \& A. Heck,
(Garching, ESO publications) p. 177}

\rref{Hirayama, T., \etal, 1995, to be submitted to PASJ}

 \rref{Johnston S., Lyne A.G., Manchester R.N., Kniffen D.A., D'Amico N.,
      Lim J., Ashworth M., 1992a, MNRAS 255, 401}

  \rref{Johnston S., Manchester R.N., Lyne A.G., Bailes M., Kaspi V.M.,
      Qiao G., D'Amico N., 1992b, ApJ, 387, L37}

  \rref{ Johnston S., Manchester R.N., Lyne A.G., Nicastro L.,
      Spyromilio J., 1994, MNRAS, 268, 430}

\rref{Kaspi, V., \etal, 1994, to be submitted to ApJ}

\rref{King, A. \& Cominksy, L., 1994, ApJ, 435, 411 (KC94)}

\rref{Kochanek, C., 1993, ApJ, 406, 638}

\rref{Koyama, K, 1988,  in {\it Physics of Neutron Stars and Black Holes},
ed. Y. Tanaka (Tokyo: Universal Academy Press), p. 55-65}

\rref{Leahy, D.A., \etal, 1983, ApJ, 266, 160}

\rref{Tavani, M., 1994,
in {\sl The Gamma-Ray Sky with COMPTON GRO and SIGMA},
eds. M. Signore, P. Salati \& G. Vedrenne
(Dordrecht: Kluwer Academics), in press}

\rref{ Tavani M., Arons J., Kaspi V.M., 1994a, ApJ, 433, L37  (TAK94)}

\rref{Tavani, M., et al., 1994b, submitted to Nature}

\rref{ Taylor J.H., Cordes J.M., 1993, ApJ, 411, 674}

\rref{Waters, L.B.F.M., \etal, 1988, A\&A, 198 200}

\rref{ Zimmermann H.U., Belloni T., Izzo C., \etal, 1993, MPE report 244}

\end{document}